\documentstyle[preprint,revtex]{aps}
\begin{document}
\begin{title}
{\bf  Scaling of  Island Growth in Pb Overlayers on Cu(001)}
\end{title}
\author{Wei Li$^{(1),(a)}$, Gianfranco Vidali$^{(1)}$
and Ofer Biham$^{(1),(2)}$}
\begin{instit}
$^{(1)}$Department of Physics,
Syracuse University,
Syracuse, NY 13244-1130
\end{instit}
\begin{instit}
$^{(2)}$Laboratory of Atomic and Solid State Physics,
Cornell University,
Ithaca, NY 14853
\end{instit}

\begin{abstract}
The growth and ordering of a
Pb layer deposited on Cu(001) at 150 K
has been studied
using atom beam scattering.
At low coverage, ordered Pb islands with
\mbox{($\sqrt{61}\times\sqrt{61}$)Rtan$^{-1}\left(\frac{5}{6}\right)$}
symmetry are formed.
This is a high order commensurate phase with 30 atoms in the unit cell.
{}From the
measurement of the island diffraction peak
profiles we find a power law
for the mean island - size versus coverage
with an exponent
$n=0.54 \pm 0.03$.
A scaling behavior of growth is
confirmed and a simple model describing island growth
is presented.
Due to the high degeneracy of the monolayer phase, different
islands do not diffract coherently. Therefore, when islands
merge they still diffract as separate islands and coalescence
effects are thus negligible.
{}From the result for $n$ we conclude that the island density is
approximately a constant in the coverage range
$0.1 < \Theta <  0.5$
where the ordered
islands are observed.
We thus conclude that most islands nucleate at
$\Theta < 0.1$ and then grow
in an approximately self similar fashion
as $\Theta$ increases.

\end{abstract}


\section{INTRODUCTION}
The study of growth and ordering of overlayers deposited
on cold single crystal
substrates poses special challenges. Recent experimental
\cite{kunkel,li}
and
computational work
\cite{sanders}
on growth of metal overlayers
on metal substrates has
shown that our understanding of the interplay of kinetic,
dynamic and energetic
processes during growth is inadequate.
Useful insight may be obtained from studies of the dynamics of
nucleation, growth and coarsening in first order phase transitions
of bulk systems
\cite{gunton,hohenberg}.
In these studies, a system in thermal equilibrium is quenched into
a nonequilibrium state in which two different phases A and B coexist.
As a result the system develops spatial inhomogeneities and finally separates
into domains of phase A and domains of phase B.
The domains of the minority
phase typically form isolated droplets inside the majority phase.
According to Lifshits and Slyozov
\cite{lifshits},
in systems with a constant order parameter
the average droplet size $\overline{R}$ increases
as a function of time t
like
\begin{equation}
\overline{R} \sim t^{x},
\label{dynexp}
\end{equation}
and $x=1/3$ during
the late
stages of the growth process.
During this process large droplets tend to grow while small droplets
evaporate and shrink.
The atoms propagate between droplets by diffusion.
Similar phenomena on surfaces have recently been studied in the context of
adsorbed monolayers
\cite{tringides,ernst}.
In these experiments it was found that
$x=1/3$ in agreement with the theory.
In other dynamical phenomena,
which occur in systems with a nonconserved
order parameter, the growth exponent is $x=1/2$
\cite{gunton,hohenberg}.

Related studies of growth have been done for
liquid droplets deposited on a surface at a constant rate
\cite{family,meakin}.
It was found that the growth is dominated by two processes.
One is the creation of new droplets, while the other is the coalescence
of two or more droplets into one larger droplet.
The droplets generally extend into the
third dimension, and can be
approximated as having a spherical shape
independent of the size.
As a result, when two droplets coalesce into one, the
surface area it covers is smaller than the sum of the areas covered
by the two droplets.
The liquid density on the surface is given by
$\rho = r \cdot t$
where $r$ is the deposition rate and $t$ is the time.
As the deposition proceeds the mean droplet radius increases
according to
$\overline{R} \sim \rho^n$.
However, in the limit of
completely flat droplets
which
do not extend into the third
dimension,
the power law is modified.
This is due to the fact that for flat droplets,
the area occupied by a droplet created after coalescence is equal to
the sum of the areas of its components, and not smaller.
As a result,
the mean droplet size increases faster than a power
law of $\rho$.
This result also applies to monolayer islands
with no internal order. However, as we later show, it does not apply for
ordered monolayer phases with high degeneracy (and a large unit cell).

The scaling of island growth
in surface adsorption for submonolayer coverage
has been
studied theoretically and numerically by Bartlet,
Evans and Tringides
\cite{evans1,evans2}.
They used a simple lattice model in which atoms are deposited at
a constant rate $r$ and then hop between empty sites as random walkers
at a constant rate $h$. An island is created when an atom reaches a site
adjacent to another atom.
They then nucleate into a stable island of
size two.
These islands grow when more atoms reach sites adjacent to them
and thus aggregate. Each island in this model, no matter how
many atoms it contains,
occupies a single site on the lattice.
Therefore, this model applies only to
the limit of low coverage, where islands are small and isolated.
Using numerical simulations of the model and a rate equation which
is derived from it Bartlet, Evans and Tringides
\cite{evans1,evans2}
find that
\begin{equation}
\overline{S} \sim \Theta^{2/3} \cdot D^{1/3}
\label{evscl}
\end{equation}
where $\overline{S}$ is the average number of atoms in an island
and $D = h/r$ is the ratio between hopping rate and deposition
rate.
If we identify the linear size of an island as
$\overline{R} = (\overline{S})^{1/2}$,
this result is equivalent to
\begin{equation}
\overline{R} \sim \Theta^{n} \cdot D^{\chi},
\label{evscl2}
\end{equation}
where
$n=1/3$ and $\chi=1/6$.
The dependence of the average island size on $D$ turns
out to be in agreement with the result obtained in Ref.
\cite{villain}
from a
completely different approach.
The simulations in
\cite{evans1,evans2}
were done for low coverage of up to 0.15 monolayer (ML).
This model
does not describe the regime of
higher coverage,
where a large fraction of the atoms are deposited directly on top
of existing islands
which cover a significant portion of the substrate area.
Also, coalescence effects are not included in the model.
The experimental results in the present paper
were obtained for higher coverage,
outside the regime studied in
\cite{evans1,evans2}.
Experimental studies of the very low coverage limit are
difficult due to the fact that the islands are very small and the
diffraction signal is too weak.

The growth of ordered islands
of Ag on Si(111) was studied by
Zuo and Wendelken
using high resolution low energy electron diffraction
(HRLEED)
\cite{zuo}.
They found
that the size distribution of islands is self-similar.
{}From the analysis of peak profiles they found that the mean linear size
$\overline{R}$ of islands exhibits a power law dependence on the coverage
$\Theta$ of the form
\begin{equation}
\overline{R} \sim \Theta^n,
\label{radius}
\end{equation}
where $n$ changes between $0.2$ to $0.35$ as the substrate
temperature is varied from 350 $^{\circ}$C to 450 $^{\circ}$C.

In this paper we present a study of island growth in Pb on Cu(001)
at 150 K.
This system was chosen since it has very intriguing
characteristics.
First, there is
a considerable lattice mismatch
(the bulk lattice constant
of Cu is
3.6 $\AA$,
compared to
4.5 $\AA$
for Pb).
Second, although for deposition at high temperature
($\sim$ 400 K)
Pb orders in submonolayer structures where most of
the atoms are in registry with the
substrate, at low deposition temperature ($\sim$ 150 K)
completely  different phases are obtained.
At low coverage we observe islands of the
\mbox{($\sqrt{61}\times\sqrt{61}$)Rtan$^{-1}\left(\frac{5}{6}\right)$}
phase.
This is a high order commensurate phase
with 30 atoms in the unit cell.
The supercell is square, while the proposed internal structure
is nearly hexagonal.
This phase results from frustration effects,
between the adlayer that typically prefers the hexagonal
symmetry (which has the largest coordination number)
and the underlying square Cu(001) substrate
which tends to induce the square symmetry.
Near one layer coverage the
\mbox{($\sqrt{61}\times\sqrt{61}$)Rtan$^{-1}\left(\frac{5}{6}\right)$}
phase disorders and is replaced by
a denser monolayer phase
with symmetry
(5$\times$5)Rtan$^{-1}\left(\frac{3}{4}\right)$,
which is a square phase with 16 atoms in the unit cell
\cite{appsur}.

{}From the analysis of the diffraction peak profile we find
that
for slow deposition rate, islands of an ordered phase
are formed and grow according to
(\ref{radius}) with
$n=0.54 \pm 0.03$.
Note that this exponent
$n$
is different from the
dynamical exponent
$x$
of Eq.
(\ref{dynexp}),
which is obtained as a function of time, for a fixed coverage.
We propose a growth model consistent with these
results.
We find that due to the large degeneracy,
coalescence effects are negligible,
while from the measured exponents we conclude that the island density
does not change within the range of coverage that we have explored.
The paper is organized as follows.
The experimental set-up is described
in Section 2.
In Section 3 we describe how the Pb layer
is characterized and
present the results of atom beam scattering experiments.
These results are discussed and
interpreted in Section 4, and summarized in Section 5.

\section{EXPERIMENTAL}

The apparatus consists of a helium beam line which is coupled to
an ultra high vacuum
(UHV)
chamber
\cite{vidali}.
We used a supersonic helium beam of 18.4 meV
incident energy,
with $\delta$v/v $\sim$ 1$\%$ velocity resolution with the beam
source at 85 K and 1200 psi helium gas pressure.
In the UHV chamber there are located: a liquid nitrogen cooled
Knudsen evaporation source, a 4-grid
low energy electron diffraction
(LEED)-Auger optics and a helium beam
detector, which is a differentially pumped quadrupole mass spectrometer with an
aperture of 0.5$^{\circ}$. This spectrometer can be rotated around the axis of
the sample manipulator and can be positioned to measure the reflected helium
beam during Pb deposition
\cite{vidali}.

The sample is fixed on a VG long-travel manipulator with X,Y, and Z
translation, and polar and azimuthal rotation.  The thoroughly desulfurized
copper sample was cleaned prior to each run by Ar
ion sputtering first at room temperature and then at 350$^{\circ}C$.
The sample was then
annealed at 580$^{\circ}$C and slowly cooled.
Helium beam scattering was
used to assess the quality of the prepared surface as in previous
studies
\cite{li,vidali}.

Pb of 99.999\% purity was deposited from a liquid nitrogen shielded Knudsen
evaporation source.  The deposition time for one monolayer of Pb was
typically about 38 minutes.
The exposure calibration was obtained from Auger
Pb and Cu signals
and from the  analysis of
ABS ``adsorption curves'' (Fig. 1).
Somewhat
faster deposition rates did not affect the results.
Although the layers deposited at 150 K were found to be metastable (see below),
there was no indication from atom beam scattering
(ABS), LEED or Auger of a rearrangement within the
layer during measurements.

There is no indication that Pb atoms are incorporated in the
Cu crystal. The original Cu(001) surface could be recovered by
heating the sample above 580 $^{\circ}$C until no trace of Pb was left
as determined by Auger electron spectroscopy. Furthermore,
Pb is virtually immiscible with Cu
\cite{immisc}.

\section{RESULTS}

\subsection{Characterization of the Pb Overlayer}

In a previous study
\cite{appsur},
we have used ABS to monitor
Pb growth on Cu(001) at low temperature.
Here we will describe in more detail how ABS data are used to
characterize the growth of a Pb layer.
Specifically, we show how the ``ABS adsorption curve'' (Fig. 1)
and the ``rocking curve'' (Fig. 2) can be used to obtain
precise coverage calibration,
the average height of Pb adatoms above Cu(001) and the
in-phase/out-of-phase scattering conditions.

We fit
the ABS adsorption curve (Fig. 1) to the
equation
\cite{poelsema}:
\begin{equation}
\frac{I}{I_0}=(1-m\Theta)^{\frac{\Sigma_{Pb}n_s}{m}}
+(m\Theta)^{\frac{\Sigma_{v}n_s}{m}}+2\left(\frac{A_1}{A_0}\right)cos\varphi
(1-m\Theta)^{\frac{\Sigma_{Pb}n_s}{2m}}(m\Theta)^{\frac{\Sigma_{v}n_s}{2m}}.
\label{adsfit}
\end{equation}
The first and the second terms represent scattering
from the uncovered parts of
the surface and from the adsorbate covered parts, respectively.
The last term
describes the interference between these contributions.
I$_0$ is the specular peak intensity of
the clean surface.
The coefficients $A_0$ and $A_1$
represent the specular scattering amplitudes:
A$_0$ is from the clean surface while
A$_1$ is the  specular peak amplitude at one monolayer completion coverage.
The parameter
m=$\frac{1}{0.64}$ is the inverse
of the ideal coverage for the monolayer
completion phase, which is
$\Theta= 0.64$
for
(5$\times$5)Rtan$^{-1}\left(\frac{3}{4}\right)$
phase (see Fig. \ref{diag}).
The coverage $\Theta$ is defined as the ratio between the number of
Pb adlayer atoms and the number of Cu substrate atoms.
Thus Fig. 1 can be used to calibrate exposure into coverages.
The quantities
$\Sigma_{Pb}$ and $\Sigma_v$ are cross sections of
the He atom scattering from isolated
Pb atoms and
vacancies in the monolayer, respectively.
Fig. \ref{ads} shows the fitting result (solid line) to
the ABS adsorption curve up to one monolayer coverage.
{}From fitting we have
$\Sigma_{Pb}$=58$\AA^2$
and
$\Sigma_v$=23$\AA^2$,
respectively.
The $\Sigma_{Pb}$ value obtained here is
slightly smaller than the 62$\AA^2$ obtained from fitting to the initial slope
of the ABS adsorption curve. The fitted interference factor,
$cos\varphi=0.99$, gives the
phase difference $\varphi$ between beams reflected
from uncovered surface and the
adsorbate overlayer.
It
is related to the overlayer height by
\cite{hg-prb}:
\begin{equation}
\varphi=2hkcos\theta_i+2n\pi, \hspace{1.5cm}(n, integer).
\label{phase}
\end{equation}
Here $h$ is the overlayer height,
$\theta_i$ is the incident beam angle and $k$ is the beam wave
vector which can be obtained by measuring the beam temperature.
We obtain that the Pb overlayer height is 2.0$\pm0.1\AA$ above
the Cu plane.

Alternatively, we can
obtain the Pb overlayer height by
taking a rocking curve,
i.e.,
ABS specular peak intensity vs. incident angle.
Fig. 2 shows such a rocking curve taken with
half layer of
the
\mbox{($\sqrt{61}\times\sqrt{61}$)Rtan$^{-1}\left(\frac{5}{6}\right)$}
structure
(the full layer coverage is 0.49 Pb/Cu atom ratio).
The intensity oscillation in Fig. 2 is due to
the interference of the scattering from the substrate and Pb layer.
Prior to Pb deposition a similar measurement was taken for the clean
Cu(001) surface and no oscillations were detected.
We have maxima at
the in-phase condition and minima at out-of-phase condition.
The following
formula
\begin{equation}
2 h k \cos \theta_i = 2 n \pi
\end{equation}
where n is an integer, gives the in-phase condition.
For half integer $n$ we have the out-of-phase condition.
By analyzing the data in Fig. 2 we obtain that the Pb layer is
$2.1 \pm 0.1 \AA$ above the Cu(001) surface.
This value is
in good agreement with the
one obtained above by fitting the ABS
absorption curve.
The above results demonstrate the equivalence of these two methods.
Actually, one can see that they are
all based on the interference between specular beams: the reflection from
the bare surface and from the deposited overlayer.
We have used this information to select the in-phase
and out-of-phase conditions for the measurements
described in Section C.

\subsection{Low Temperature Phases}

In Fig. \ref{diag}  we present a sketch of the phase diagram. The two phases
below 0$^{\circ}$C have not been reported by other groups before
\cite{sepulveda,ferrer}.
We have presented the
(5$\times$5)Rtan$^{-1}\left(\frac{3}{4}\right)$ phase structure elsewhere
\cite{li,appsur}.
Here we concentrate
on the submonolayer
($\sqrt{61}\times\sqrt{61}$)Rtan$^{-1}\left(\frac{5}{6}\right)$ phase.
This phase, as we show in Fig. \ref{diag},
starts to appear at very low coverage
around $\Theta=0.1$ and continues
up to about 0.5.
The LEED pattern of
($\sqrt{61}\times\sqrt{61}$)Rtan$^{-1}\left(\frac{5}{6}\right)$ turns out to be
quite similar to that of (5$\times$5)Rtan$^{-1}\left(\frac{3}{4}\right)$
\cite{appsur},
and consists of contributions from two types of reciprocal nets.
In Fig. \ref{stru} the reciprocal lattice nets and proposed structure of
($\sqrt{61}\times\sqrt{61}$)Rtan$^{-1}\left(\frac{5}{6}\right)$ are presented.
The two types of reciprocal nets are from two kinds of real space domains as
illustrated by two big squares in Fig. \ref{stru}(b),
and are rotated by
tan$^{-1}$(5/6) and tan$^{-1}$(6/5) from the Cu $<1 \bar{1} 0>$ direction.
The
LEED spots
actually observed are indicated in Fig. \ref{stru}(a) with dark points, and
many of these are from two domains and too close to be resolved.   Since
the observed LEED pattern is
rather faint, the ABS
diffraction from this phase (Fig. \ref{scan})
has played an important role to
determine the superstructure.
Within the perimeter of the
superstructure we propose a compact pseudo-hexagonal arrangement
with an ideal coverage of 0.49.
A similar kind of structure has been proposed
before for Bi on Cu(001)
\cite{rhead}.

The two low temperature ordered
phases are obtained after deposition at 150K  and are
likely to be metastable.
Upon heating them above about 0$^{\circ}$C they convert
into the phases obtained by depositing Pb at 400 K,
and they cannot be reached by cooling the overlayer
deposited at room temperature.
These
high temperature phases
are equilibrium phases, since
they can be melted and recrystallized
\cite{li,ferrer}.
The low temperature phases
are high order commensurate structures with
only one atom in 16 for
the (5$\times$5)Rtan$^{-1}\left(\frac{3}{4}\right)$
or three atoms in  30 for the
($\sqrt{61}\times\sqrt{61}$)Rtan$^{-1}\left(\frac{5}{6}\right)$
in registry with the substrate.
This suggests
that they are the result of limited diffusion and
of a delicate competition between Pb-Pb and Pb-Cu interactions.
It is unusual to find commensurate
structures with so few atoms in registry with the substrate; in fact,
in most cases the overlayer structure becomes incommensurate with a hexagonal
unit cell.
A general procedure for classification of these phases has been
proposed
\cite{nine-2},
while
a calculation of their energetics is in progress
\cite{tobepub}.

\subsection{Self-similar growth of Pb islands}

As one can see from Fig. \ref{diag}, the
($\sqrt{61}\times\sqrt{61}$)Rtan$^{-1}\left(\frac{5}{6}\right)$ phase exists at
coverages well below its ideal coverage of 0.49, which is an indication of
island growth.
In Fig. \ref{spe-pro}, specular peak profiles
taken in the nearly in-phase condition are plotted
at a few
representative Pb coverages.
Basically, each peak is composed of
two parts: one is the sharp and narrow top peak which is
reflection from uncovered Cu
substrate; and  the other is the broaden and shoulder-like
tail part which is due
to reflection from Pb islands.
In the out-of-phase condition, only a broad peak is observed.
Although it is possible to extract information
on island growth from this measurement,
in reality the analysis is complicated by the fact that
the ABS specular peak has contributions
both from the substrate and islands.
In this instance it is far more advantageous
to use the
ABS diffraction peak
due to ordering within
Pb islands to investigate how islands grow
since there is no interference from the substrate.
Fig.
\ref{dif-pro}
shows the evolution of the profiles of (0,-1) peak
\cite{peaklbl}.
The
decreasing of the peak width with the increasing of the Pb coverage indicates
the growth of islands.

The measured
peak profiles
$I_M(\vec k_{\parallel})$
are actually convolutions of the instrument response
function
$T(\vec k_{\parallel})$
with the "true" peak shape
$I_s(\vec k_{\parallel})$
of the system we
are investigating.
Two procedures have been applied to extract "true" peak shapes
from:
\begin{equation}
I_M(\vec k_{\parallel}) = \int T(\vec k_{\parallel} - \vec S_{\parallel})
I_s(\vec S_{\parallel}) d \vec S_{\parallel}.
\label{convo}
\end{equation}
First
we performed an analysis of the diffraction beam shapes by deconvolving the
instrument response function  using Fourier analysis.
For our ABS
system, the instrument response function can be
measured by positioning the detector facing the direct incident beam.
A small correction, to take into account velocity broadening,
has been applied when fitting diffraction peaks.
For the best prepared Cu(001) surfaces there is hardly any broadening
in the specularly reflected peak.
Essentially, what we do is to Fourier transform  the
measured peaks and the instrument response function
(which is a Gaussian).
Using the convolution theorem
\cite{sneddon}
we find that
\begin{equation}
\hat I_M(\vec k_{\parallel})
= \hat T(\vec k_{\parallel})  \cdot
  \hat I_s(\vec k_{\parallel})
\label{convol}
\end{equation}
where
$\hat I_M$, $\hat T$ and $\hat I_s$
are the Fourier transforms of
$I_M$, $T$ and $I_s$
respectively.
{}From this we obtain
$\hat I_s(\vec k_{\parallel})  =
 \hat I_M(\vec k_{\parallel}) / \hat T(\vec k_{\parallel})$
and then we
perform an inverse Fourier transform to get
"true" peak profiles
$I_s(k_{\parallel})$.
The advantage of this process is that  these deconvolved
data have been obtained without assuming any functional form for the "true"
signal.  In Fig.
\ref{conv}
we present  deconvolved peak profiles from three
representative coverages after  normalizing intensities and scaling the
horizontal axis. The oscillations in the wings of the peaks are due to the
deconvolution procedures; in fact, they are not present in the original data
(see Fig. \ref{dif-pro}).
The diffraction profiles represent the structure factor of
the physical
system.
Through our measurement, the different peak profiles
coincide with each other after  scaling as shown in
Fig. \ref{conv}. This indicates
that the structure factor for our system is a scaling function and is
independent of Pb coverages
\cite{zuo-jvst}.
We conclude that growth of
Pb islands
is self-similar in the range of coverage investigated.

Comparing
Fig. \ref{conv}
with
Fig. \ref{dif-pro}
one can see that  some additional noise is introduced in
the deconvolved data due to the deconvolution procedure.
Using this method it is
hard to determine the
analytical form of the "true" peak shape from the curves in Fig. \ref{conv},
especially when there is a considerable amount of noise.
We used another procedure to analyze peak profiles analytically.
The
measured peak profiles are fitted directly by the convolution of
the instrument response function (Gaussian) with a chosen function, which is
supposed to be the "true" diffraction peak shape.
Following analyses of island growth studied by HRLEED, we tried Gaussian,
Lorentzian and power Lorentzian as fitting functions
\cite{vidali,li2}.
In Fig. \ref{dif-pro}, fits using a power Lorentzian:
\begin{equation}
I(\Delta K_{\parallel}) \propto
\frac{A}{(\xi^{2}+\Delta K_{\parallel}^2)^m}
\label{eq:lor}
\end{equation}
\noindent
with m=5 are presented for three different coverages. The same fitting function
was used for all coverages, from about
$\Theta = 0.1$ to $0.5$.
The fact that an
identical function fits well all the line shapes is again a confirmation of the
scaling behavior.
However, since there is no theoretical reason to use Eq.
(\ref{eq:lor}),
it should be considered at this stage only as a
convenient fitting function.

The inverse
width of the deconvolved peak (FWHM of Eq. (\ref{eq:lor})) is displayed as a
function of coverage in Fig.
\ref{fwhm}.
The growth of the mean island size
$\overline{R}$ versus coverage can be well described by the equation
\cite{zuo}:
\begin{equation}
\overline{R} \sim \frac{1}{FWHM} \propto \Theta ^n
\label{eq:cov}
\end{equation}
We find that the best fit
for the data in Fig. \ref{fwhm} is obtained with
$n=0.54 \pm 0.03$.
We estimate that the
average size of ordered islands goes from
about 30$\AA$ to 100$\AA$. For the
case of growth of two-dimensional islands, we are not aware of any theory that
makes a prediction about the exponent in Eq. (\ref{eq:cov}).
The value of the exponent  we obtain is
different from the experiment of Zuo and Wendelken
\cite{zuo}
where $n$ was found
to vary between
0.2 at $T=340^{\circ}C$
and
0.35 at $T=450^{\circ}C$.
The difference between these two results are discussed in the next
Section.
The temperature dependence of $n$ might indicate
that different processes are
present at different temperatures.

Fig. \ref{inten} presents our result of the (0,-1) peak intensity as a function
of Pb exposure plotted in a ln-ln scale.  We can establish a power law by
fitting the data in Fig. \ref{inten} to:
\begin{equation}
I_{01}\sim\Theta^{\tilde{p}}.
\label{p-value}
\end{equation}
The solid line in Fig.
\ref{inten}
is the best-fit result,
 with
$\tilde{p}= 1.89 \pm 0.04$.

Note that after a given amount of Pb is deposited we find that
the diffraction on specular intensities didn't change during data taking
(from a few minutes to tens of minutes).
This is indicating that the layer has
stopped evolving before measurements are taken.

\section{Discussion and Interpretation}

To understand the scaling behavior of the
islands we first define the
island density $N(\Theta)$
which is expected to exhibit a
power law dependence on the coverage:
\begin{equation}
N(\Theta) \sim \Theta^q.
\label{density}
\end{equation}
Since the coverage provides the average density of
adlayer atoms, which are distributed between islands of various
sizes
\begin{equation}
\Theta \sim N(\Theta) \cdot \overline{S}(\Theta)
\label{powers}
\end{equation}
where
$S(\Theta) = \overline{(R^2)}$
is the average island area.
Using the scaling assumption we can write
\begin{equation}
P(R(\Theta), \Theta)
= { 1 \over { \left( \overline{R(\Theta)} \right)^{\lambda}} }
P^{\prime}(x)
\label{scalin}
\end{equation}
where
$P(R(\Theta), \Theta)$
is the probability, at coverage $\Theta$,
of finding an island of size $R$,
$P^{\prime}(x)$
is a scaling function independent of $\Theta$ and
$x = R / \overline{R}$.
As a result, we obtain
$\overline{(R^2)} = ({\overline{R}})^2$.
In this case one can replace equation
(\ref{powers})
by
\begin{equation}
\Theta \sim N(\Theta) \cdot ({\overline{R}})^2 \sim \Theta^{q+2n}.
\label{powers2}
\end{equation}
The scaling assumption thus leads to the relation
\cite{zuo}
\begin{equation}
q+2n =1.
\label{qnrule}
\end{equation}

Submonolayer island growth generally
involves two processes in addition to the growth of existing islands:
the creation of new islands
by nucleation of atoms
as the coverage increases
and the coalescence of two or more islands into one larger island.
The first process tends to increase the island density while the
second process decreases it.
Typically, in the early stages of growth the islands are very small
and isolated, and coalescence is rare.
On the other hand, during later growth stages, at relatively high
temperature and low deposition rates, very few new islands form,
since mobility is high enough for atoms to
aggregate into existing islands.
We can thus use Eq.
(\ref{qnrule})
to identify two general regimes of $q$ and $n$.
In systems where islands do not coalesce,
their density can only increase, and therefore
$q \geq 0$ and $n \leq 1/2$.
On the other hand in cases where islands coalesce but no
new islands appear,
$q \leq 0$ and $n \geq 1/2$.
In systems where both processes occur simultaneously, the values of
$q$ and $n$ will be determined by the balance between them.

We will now show that due to the structure of the
($\sqrt{61}\times\sqrt{61}$)Rtan$^{-1}\left(\frac{5}{6}\right)$
phase,
coalescence is practically negligible in our system.
The phase
($\sqrt{61}\times\sqrt{61}$)Rtan$^{-1}\left(\frac{5}{6}\right)$
has a large unit cell with 30 atoms, only three of them are in perfect
registry with the substrate. It thus exhibits a high degeneracy
due to both rotations and translations.
The supercell
can appear with two rotation angles with respect
to the substrate:
$\theta = tan^{-1}(5/6)$
or
$\theta = tan^{-1}(6/5)$
(see Fig. 4)
and therefore is doubly degenerate.
In addition the internal structure of the unit cell can appear
in two degenerate states rotated by $90^{\circ}$ with respect to
each other.
The translational degeneracy is due to the large unit cell and the
fact that only three atoms are in
perfect registry with the substrate.
Careful analysis of the proposed unit cell structure, which
covers 61 lattice sites of the substrate shows that
there are 21 translationally degenerate state.
We thus conclude that the total degeneracy in the
system is 84-fold.
The high degeneracy has a substantial effect
on the island growth.
When two islands start growing into each other and merge it is
very unlikely that they will match properly.
Since they are degenerate and not in phase they
will not diffract coherently.
As a result, for our diffraction experiments they behave like
separate islands even if physically connected.
Therefore, there is practically no coalescence in the system as
far as diffraction measurements are concerned.
The only process that may occur, in addition to the growth of
existing island, is the creation of new islands.
Therefore, the range of values available for $q$ and $n$
is $q \geq0$ and $n \leq 1/2$.

In our experiment $n = 0.54 \pm 0.03$, which  is very close to $1/2$.
{}From Eq.
(\ref{qnrule})
we obtain that
$q = -0.08 \pm 0.06$,
which is close to zero, but negative
and may indicate a slight decrease in the island density between
$\Theta=0.1$
and
$\Theta=0.5$.
Since we know that there is practically no coalescence in our system we
conclude that as the coverage increases
between
$0.1<\Theta<0.5$
no new islands are created and it is thus very close to the marginal case of
$q=0$ and
$n=1/2$.
{}From the relation
$\overline{R} \sim \Theta^n$
we obtain
\begin{equation}
{d \overline{R} \over d \Theta} \sim \Theta^{n-1} \sim
{\overline{R}}^{(n-1) / n}
\end{equation}
for the average radius and
\begin{equation}
{d \overline{S} \over d \Theta} \sim  \Theta^{2 n - 1}
        \sim {\overline{S}}^{(2n-1) / 2n}
\end{equation}
for the average area.
For $n=1/2$ we find that $d \overline{S}/d\Theta = C$
and $C$ is a constant.
This is consistent with a growth model in which the islands are
randomly distributed on the surface and each island has an area
around it from which it collects the adsorbed atoms.

According to this picture the islands are created in the limit of low
coverage of $\Theta < 0.1$.
The dominant mechanism for the creation of new islands
seems to occur
when two or more atoms meet and nucleate together after deposition
\cite{evans1,evans2}.
Additional islands may appear when atoms nucleate on defects,
steps or impurities on the surface.
However, these
can account for a small number of islands.

A simple model that may describe the growth process in this system
is based on the Voronoi construction
\cite{weaire}.
In this construction one first defines a set of random points,
or centers, on the surface.
One then draws the perpendicular bisecting lines to the lines
joining any two centers. The smallest convex polygon around each
center will contain all the points which are closest to
this center.
In this model one assumes that a random distribution of islands is
initially created at very low coverage.
As more atoms are deposited, they diffuse and aggregate into existing
islands which then grow.
Assuming that each atom tends to aggregate into the nearest island,
each island attracts the atoms that fall in
the Voronoi polygon around it.
The size of each island will thus be proportional to the size of
the Voronoi polygon around it, at all stages of the growth process.
Therefore, within the assumptions of this model, the growth will
be self similar.
In reality things are more complicated. The initial seeds of islands
may not be completely random. Not all deposited atoms aggregate into
the nearest island, and most importantly, at larger coverage the
important distance is between the deposited atom and the boundary of
the island rather than to its initial seed. However, we believe
that the island size is proportional to the size of its Voronoi
polygon even for high coverage.

This model may also provide a clue for why no new islands seem to
appear above some low coverage.
A rough estimate
shows that an island which nucleates at
$\Theta = 0.18$ may reach a size which is only about $10 \%$
of an island created at $\Theta = 0$
\cite{tobepub}.
This results from two reasons:
first, at $\Theta = 0.18$ the new island has negligibly small
size, compared to
the finite existing islands.
Then, the domain from which it attracts more atoms is much smaller
since it should be drawn by bisecting the distance between its
seed and the boundaries of the existing islands.
Therefore, it will grow much more slowly than islands which were
created earlier.
This is a particularly important effect in our system which has
a very large unit cell.
Islands that nucleate late may not reach the minimal size
needed for diffraction.
Islands that nucleate in the late stages of the deposition process
thus seem to be "shadowed" by the existing islands.
We plan to explore these ideas both analytically and using
computer simulations in order to obtain quantitative predictions
\cite{tobepub}.

Zuo and Wendelken
observed that the diffraction peak intensity in their experiment
increases like $I \sim \Theta^p$ and also
obtained a relation between the exponents
$n$ and $p$
with a Gamma domain-size distribution.
As long as the scaling growth exists,
the following relation should be satisfied
\cite{zuo}:
\begin{equation}
p=1+2n.
\label{relation}
\end{equation}
In our experiment we use a linear slit detector which, in fact,
integrates over one dimension in $k$ space
while the scan is done over the perpendicular direction.
Using our power Lorentzian fits we find that the FWHM of our
integrated peak scales with
$\Theta$ exactly like the original peak. However, the integrated peak
intensity $I_{01}(k_x)$ scales like the product of the original
peak intensity $I_{01}(k_x,k_y)$ and its width (FWHM).
Since the width scales like $\Theta^{-n}$
we conclude that $\tilde{p} = p - n = 1 + n$.
Our result, $\tilde{p}=1.89 \pm 0.04$
is much larger than $1+n = 1.54 \pm 0.03$.
This may be due to the fact that the slit detector is
not very sensitive near its ends and thus does not provide a
complete integration.

\section {Summary}

Using ABS, we have studied Pb growth
on Cu(001) substrate
at 150 K.
Two high order commensurate
phases have been discovered.
The
($\sqrt{61}\times\sqrt{61}$)Rtan$^{-1}\left(\frac{5}{6}\right)$
phase has a square unit cell with a proposed quasi-hexagonal
internal structure.
This structure seems to be energetically favored since it provides a
compromise between the adlayer-adlayer interactions which favor the
hexagonal structure and the substrate that tends to induce the
square symmetry.
We
found
that the Pb layer grows in
($\sqrt{61}\times\sqrt{61}$)Rtan$^{-1}\left(\frac{5}{6}\right)$
islands before forming the
(5$\times$5)Rtan$^{-1}\left(\frac{3}{4}\right)$
phase which is a square phase.
By analyzing line shapes, we are able to show that island growth is
self-similar.  Power growth laws for mean island size and diffraction peak
intensity are established.
Due to the high degeneracy of the monolayer phase, when islands merge
they form a boundary line. They also do not diffract coherently
and therefore coalescence effects are negligible in our
diffraction experiments.
{}From the experimental results we conclude that in this particular
system, ordered islands tend to form at very low coverage and
then to grow in an approximately self similar fashion as the
coverage increases.
In the future we plan
to explore the temperature dependence of the exponents
$n$ and $\tilde{p}$, and to examine how the initial island
density is determined.

\section {Acknowledgements}

This work was supported by NSF grants DMR-9119735 (G.V)
and DMR- 9217284 (O.B).
The work at Cornell (O.B) was supported by NSF grants
DMR-9118065 and DMR-9012974.
We thank J.-S. Lin and H. Zeng for technical assistance
and B. Cooper, T. Curcic, J. Evans, V. Elser,  J. Sethna
and M. Tringides for helpful discussions.
\vspace{0.5in}

\noindent
$^{(a)}$ Present address: The James Franck Institute, University of
Chicago, Chicago, IL 60637.

\figure{Pb low temperature ABS adsorption curve: I(0,0) vs. Pb coverage.
Surface at 150 K, $\theta_i$=60 $^{\circ}$. Solid line is a fit to
Eq. (\ref{adsfit}).
\label{ads}}

\figure{ABS specular peak intensity vs. incident angle,
half layer Pb deposited.
The substrate temperature is $T_s = 150 K$.
\label{absspec} }

\figure{Sketch of phase diagram of Pb on Cu(001) obtained by ABS and LEED
data. Crossed region: disordered; cross-hatched region: phases
coexist.
\label{diag}}

\figure{(a) Reciprocal space: solid and dashed lines are from two domains; (b)
proposed real space structure of
($\sqrt{61}\times\sqrt{61}$)
Rtan$^{-1}\left(\frac{5}{6}\right)$.
The unit cell which is square
contains 30 atoms
which are arranged in a
pseudo hexagonal structure.
\label{stru}}

\figure{ABS diffraction scan from
($\sqrt{61}\times\sqrt{61}$)Rtan$^{-1}\left(\frac{5}{6}\right)$,
$\theta_i$=60$^{\circ}$
for full coverage of
$\Theta = 0.49$.
The substrate temperature is $T_s = 150 K$.
We identify the labling of the peaks as
(A) $(36/61,30/61)$, $(36/61,31/61)$;
(B) $(30/61,25/61)$, $(31/61,25/61)$;
(C) $(24/61,20/61)$;
(D) $(18/61,15/61)$;
(E) $(12/61,10/61)$; and
(F) $(6/61,5/61)$, $(5/61,6/61)$.
Note that peaks A, B and F are double peaks due to overlap between
peaks of the two grids.
\label{scan}}

\figure{ABS specular peak profiles at surface temperature of 150 K,
$\theta_i$=60$^{\circ}$
and different coverages.
Notice the emergence of wings indicating island growth.
\label{spe-pro}}

\figure{(0,-1) ABS diffraction peaks from
($\sqrt{61}\times\sqrt{61}$)Rtan$^{-1}\left(\frac{5}{6}\right)$ islands at
representative coverages; Solid lines are the fitting results of the
convolution of the instrument response function
with the function in Eq. (\ref{eq:lor}). $\theta_i$=60$^{\circ}$.
{}From the full width at half maximum (FWHM) of such peaks
we obtain the scaling of the island size as a function of coverage.
\label{dif-pro}}

\figure{ABS diffraction data from islands after deconvolution of the
instrument response function. Data from
three representative
coverages have been rescaled and plot in the same  graph.
$\theta_i$=60$^{\circ}$.  w is the full width at half maximum
(FWHM) of a peak.\label{conv}}

\figure{ln(Inverse FWHM) vs. ln(coverage)
obtained from an analysis of diffraction
peaks as in Fig. \ref{dif-pro}.
The line is the best-fit through the data, see
Eq. (\ref{eq:cov}).
The slope is
$n = 0.54 \pm 0.03$ for coverage between $0.1$ and $0.5$.
\label{fwhm}}

\figure{A ln-ln plot of the (0,-1) peak intensity vs Pb coverage.
The solid
line is the best-fit to Eq. (\ref{p-value}).
The slope is
$\tilde{p} = 1.89 \pm 0.04$.
\label{inten}}


\begin{references}
\bibitem{peaklbl}   Diffraction peaks are labeled according to Fig. 4a.
                    ABS is sensitive to the top layer only, so there
                    is no contribution from underlying Cu atoms.
                    However, there could be a contribution from clean
                    Cu(001) areas. In reality, recently
                    we observed that, when Pb is adsorbed at 400K
                    at low coverage $(\Theta \simeq 0.15)$,
                    randomly adsorbed Pb atoms produce a change in
                    the composition of the electronic charge density
                    profile of neighboring Cu atoms yielding a small
                    (0,-1) diffraction peak \cite{li2}. Without Pb adsorption
                    the Cu(001) surface is so flat that
                    no diffraction peaks are seen.
                    In the present case this effect is small and confined
                    to the smallest coverages $(\Theta < 0.1)$.
                    Thus the $(0, -1)$ diffraction peak is overwhelmingly
                    due to Pb islands. Strong fractional order peaks have
                    been observed, but were not used in the fitting because
                    they were too close to the tails of the dominant $(0,0)$
                    peak.
\end{references}
\end{document}